\documentstyle[11pt,psfig,aaspp4]{article}
\begin{document}

\title{9286 Stars: An Agglomeration of Stellar Polarization Catalogs}
\author{Carl Heiles}

\affil{Astronomy Department, University of California,
    Berkeley, CA 94720}

\altaffiltext{1}{email: cheiles@astro.berkeley.edu}

\begin{abstract}
	This is a revision.  The revisions are minor.  The new version
of the catalog should be used in preference to the old.  The most
serious error in the older version was that $\theta_diff$ was incorrect,
being sometimes far too large, for Reiz and Franco entries; the correct
values are all zero for that reference. 

	We present an agglomeration of stellar polarization catalogs
with results for 9286 stars.  We have endeavored to eliminate errors,
provide accurate ($\sim$arcsecond) positions, sensibly weight multiple
observations of the same star, and provide reasonable distances.  This
catalog is available by anonymous FTP as ascii file
ftp://vermi.berkeley.edu/pub/polcat/p14.out .  This manuscript is also
available as the postscript file
ftp://vermi.berkeley.edu/pub/polcat/pol1.ps .  \end{abstract}


\section{INTRODUCTION}

	Polarization has been measured for thousands of stars and
presented in perhaps a dozen catalogs.  Some previous attempts to
combine these lists are very admirable because they have made it much
easier to use the data.  The largest include Mathewson et al (1978;
hereafter MFKNK) catalog (CDS catalog II/34A) and Axon and Ellis (1976)
(CDS catalog II/178).  However, they have deficiencies; for example,
both list multiple results for individual stars and have not purged
errors from the original catalogs.  The present agglomeration combines
multiple observations with weighted averages, fixes most errors,
provides accurate positions, and reasonable estimates for stellar
parameters such as distance and extinction.  It also includes
information on which original catalogs were used for each entry.

          Section 3 discusses the catalogs that we have included,
together with the information contained in each.  The MFKNK, Axon and
Ellis (1976), Reiz and Franco (1998) and Goodman (1997) catalogs were
originally provided to us in electronic form.  We entered the
Appenzeller (1974) catalog by hand from the printed page.  For all the
other catalogs, we scanned printed versions and converted them to ascii
files.  However, in the final analysis, all of these catalogs contain
data that were entered into a computer file by hand.  Therefore they
contain potential errors.  Heiles (1997) recounts a few problems
regarding such errors, and one goal was to eliminate as many problems as
possible. 

\section {FINDING AND FIXING POSITION AND IDENTIFICATION ERRORS}

	Nearly all positions in our agglomeration are derived from one
of the four primary stellar databases, which are the Hipparcos, Tycho,
SAO, and SIMBAD databases.  Below we refer to these as the {\it primary
stellar databases}.  In our agglomeration, $IDCAT$ tells which of these
databases provided the position. 

        Many catalogs provide both positions and an identification
number (an HD, BD, CD, or CPD number).  However, sometimes the position
and identification number are incommensurate.  We caught these problems
by comparing the polarization catalog's star identity and position with
those from the primary stellar databases.  Comparison with Hipparcos,
Tycho, and SAO was done automatically when one or more actually listed
the catalog star, which was the case for about $98\%$ of the stars.  For
the remaining stars we examined the SIMBAD database by hand.  The
polarization catalogs usually list positions to within a few arcminutes;
our criterion for acceptance of the catalog star as listed was that the
catalog and primary stellar database positions agree to within $5.4'$. 
This tolerance may seem overly large, but we are confident that it is
reasonable on the basis of empirical examination. 

          In every case of a successful identification, we adopted the
position from the primary stellar database instead of from the
polarization catalog.  This means that our positions are accurate at the
arcsecond level.  For the automatic comparisons there is no ambiguity or
possibility for error.  For our SIMBAD identifications there is a
possibility for error in our compilation because we entered the
identification and positional information by hand.  However, if we did
make a mistake in this process, then the position can be incorrect by up
to $\sim 10'$ but no larger, because larger position discrepancies would
have been caught automatically.  For some cases of SIMBAD
identifications, the star identification name was not entered in the
appropriate field; however, the star position is correct. 

          There were a nontrivial number of unsuccessful
identifications, and because polarization data are valuable but few, so
we attempted to discover why.  For unsuccessful comparisons there are
three possibilities: the original polarization catalog contains a
simple, single typographical error; it contains multiple typographical
errors; the identification or position is simply incorrect. 

	We first attempted to reconcile all discrepancies to a simple,
single typographical error.  We assumed that if there was an error in
any {\it one} of the right ascension, declination, or star
identification, then we could reasonably assume that the error was
typographical.  This was the case for the overwhelming majority of
discrepant objects, and we made the appropriate change in the original
database; we do not flag such typographical errors in our compilation. 
If there was an error in {\it more than one} of the three parameters,
then we assumed that the error was fundamental and flagged it by setting
$IDCAT = -999$\footnote{ The one exception to this rule was several
stars in the MFKNK catalog that had large errors in declination---one
was listed with $75^\circ$ instead of $57^\circ$ and a few with the
wrong sign.  Because of the precession, mentioned below, not only the
declinations but also the 1950 right ascensions are incorrect; we did
not flag these stars with --999.}.  Some stars have no listing in any of
the four stellar position databases and they also have $IDCAT=-999$. 

	In summary, the positions or identification numbers of 
stars that have $IDCAT=-999$ are not absolutely certain, either because
the information is discrepant or because there wasn't enough information
available to check. Finally, an entry in any parameters of $-999$,
$-999.9$, or $-99$ means that the information was not available.

\section {COMMENTS ON INDIVIDUAL POLARIZATION CATALOGS}

          The following comments are not guaranteed to be complete or
accurate. The listing is in the reverse order of $POLREFS$ in Table 3.

\subsection{ The MFKNK catalog}

          The MFKNK catalog is available from the CDS catalog service as
catalog number II/74A.  It is a huge compilation of, first, the original
Mathewson and Ford (1970; MF) compilation of their own and others
measurements; and the Klare and Neckel (1978; KN) measurements.  The KN
data set is a particularly valuable addition because it contains many
stars in the third and fourth Galactic quadrants.  However, the error
rate in the KN section of the MFKNK catalog is relatively large.  This
is strange because the {\it paper} version of the KN catalog differs
from the MFKNK {\it electronic} version, and in cases of discrepancy it
is the {\it paper} version that is correct.  A few examples: --58 510 is
really HD 58510; HD 290377 is really HD 298377; --63 1513 is really --63
2513. 

          MFKNK had a systematic error: all declinations in the range
$0^\circ \rightarrow -1^\circ$ were denoted as positive instead of
negative.  MFKNK list positions for equinox 1950, but most of their star
positions are derived from lists where the equinox was 1900.  Thus MFKNK
precessed the 1900 positions forward by 50 years to get their 1950
positions, and it seems that the negative signs were missing in the
original 1900 positions too.  This means that their 1950 declinations
contain more than just a simple sign error. Positions in our agglomeration
are correct, having been taken from the stellar databases.

	All of the equatorial position angles in MFKNK for their
reference 6 (Schmidt) were set equal to zero, probably because the
original reference listed all results in Galactic coordinates.  We
assumed that the Galactic angles listed were correct. 

	Many polarizations listed in MFKNK are zero; these measurements
are really upper limits, and we did not include these incorrect results
in our final list or in our weighted averages. 

\subsection{BERDYUGIN,  SNARE, AND TEERIKORPI}

          Berdyugin et al (1995) observed 51 stars at high positive
Galactic latitudes and provided identifications, positions, and
polarization, position angle (both equatorial and Galactic), visual
magnitude, E(b--y), and distance.  Five stars had unmeasurably low
polarization and we omitted them.  For one star, BD +20 2870, the
equatorial and Galactic position angles do not agree; we assume this is
a typographical error but have no way of knowing which angle is the
correct one.  We include this star, but the large $\theta_{diff} \sim
70^\circ$ is telling.  In our compilation, which lists E(B--V) when it
is available, we have converted E(b--y) by multiplying by the factor
11/8, which is the ratio of the separation of central wavelengths of
(B--V) and (b--y). 

\subsection{KORHONEN AND REIZ}

          Korhonen and Reiz (1986) observed about 470 stars and provide
identifications, positions, and polarization, position angle (both
equatorial and Galactic), and visual magnitude.  There are two groups of
stars: their Table 1 contains 118 stars that were previously observed by
MF, and they provide detailed comparisons of the results; their Tables 2
and 3 contain 357 additional stars.  Table 1 had one identification
error and, in addition, the conversions to Galactic coordinates and to
Galactic position angle for HD 1461 were incorrect; we accept the
equatorial position angle.  Table 3 had two minor positional errors;
also, SAO 167576 had a small error in conversion to Galactic coordinates
and to Galactic position angle, which we ignore. 

\subsection{KRAUTTER}

          Krautter (1980) observed 313 stars, mostly near the Galactic
plane, and provided identifications, positions, polarization, position
angle (both equatorial and Galactic), visual magnitude, spectral type,
B--V, A$_V$, and distance.  There are a few positional errors. 


\subsection{MARKKANEN}

          Markkannen (1979) observed 31 stars and presented an
additional 41 from Appenzeller (1968; these also exist in MFKNK); he was
studying the North Galactic Polar region.  He provided identifications,
polarization, position angle, visual magnitude, spectral type, and
distance.  He did not provide positions, so we cannot be absolutely
certain about typographical errors.  Many stars had unmeasurably low
polarization and we omitted them.  There are two identification errors:
HD 110056 is really HD 111056; and we eliminated HD 114727, which must
be misidentified because it lies too far outside his area of interest. 

\subsection{SCHROEDER}

          Schroeder (1976) observed 495 stars and provided
identifications, positions, and polarization, position angle (both
equatorial and Galactic), visual magnitude, spectral type, and distance. 
For identification he provided either HD numbers or numbers from the
General Catalogue of Trigonometric Parallaxes (Jenkins 1963); for the
latter, we obtained the BD, CD, or CPD numbers.  There seems to be one
identification error: GCT 301 is the same as either CD --52 291 or CPD
--52 291, but the Schroder positions do not agree with SIMBAD's. 

\subsection{APPENZELLER}

          Appenzeller (1974) studied 126 stars in the vicinity of
Eridanus loop region and concluded that the magnetic field was deformed
in the shape of a ``magnetic pocket''.  He provided identifications,
positions, and polarization, position angle, but no spectral types or
colors.  The star he lists as HD 288553 is really HD 288353. 

\subsection{GOODMAN}

	Goodman's (1997) catalog contains stars primarily in or near
dark clouds and contains only positions and measured polarizations. 
There are no stellar identification data, so we could not perform checks
on position and we set $IDCAT=-999$ for all of Goodman's stars. 

\subsection{LEROY}

	Leroy observed about 1000 nearby stars and found zero
polarization in most of them.  However, 25 of these stars have
measurable polarization (Leroy 1993).  Leroy provided stellar
identifications, positions, polarization, position angles in both
equatorial and galactic coordinates, spectral type, and distance.  We
detected no typos in Leroy's list.  Leroy gave two distances; we used
his distances $D_i$, which are supposed to be better. 

\subsection{BEL, LAFON, AND LEROY}

	Bel et al (1993) observed stars near the Cepheus flare (near the
NCP) and provided identifications, positions, distance, polarization,
and position angle in Galactic coordinates.  There were 133 entries and
two typographical errors on positions, for S10278 and HD 678.  The
distances for many stars have alphameric suffixes such as ``mx'', and
these are unexplained; we have ignored them. 

\subsection{REIZ AND FRANCO}

	Reiz and Franco (1998) measured 361 stars that sample 35 of
Kapteyn's selected areas in the third and fourth Galactic quadrants for
$|b| \leq 30^\circ$.  This catalog appears to have accurate photometry,
reddening, and distances.  Before this paper appeared, we had finished a
preliminary version of the agglomeration in which we had specified
E(B--V) to only one decimal place.  This accuracy is insufficient for
Reiz and Franco's catalog, so we list their entries to two decimal
places.  As with Berdyugin et al (1995), we converted E(b--y) to E(B--V)
by multiplying by the factor 11/8.  This catalog provides measurements
at three wavelengths for every star.  We averaged these according to the
prescription below in section 4. 

	The Reiz/Franco catalog contained four stars that already
existed in our preliminary agglomeration: HD98310, HD98722, HD99545, and
HD100198.  For these stars we followed the easier option of choosing the
better data instead of taking a weighted average.  In two cases, HD98310
and HD100198, the older polarization errors from MFKNK were smaller so
we used the MFKNK data. 

	It is instructive to compare the distances: For the first two
stars the Reiz/Franco distances were comparable with the Neckel et al
(1980) ones.  However, for the last two the distances were widely
discrepant.  HD99545 had distances (412, 3020) pc for (Reiz/Franco,
MFKNK) and HD100198 had distances (188, 2371) pc for (Reiz/Franco,
Neckel et al).  It is difficult to know which distances are correct. 
HD99545 is a particularly difficult case because the Reiz/Franco
(polarization, reddening, distance) are about $({1 \over 2}, {1 \over
3}, {1 \over 8})$ the MFKNK values; this makes the set of parameters
reasonably compatible for {\it both} Reiz/Franco and MFKNK.  We chose
the Reiz/Franco polarization, reddening, and distance.  The case of
HD100198 is much clearer: the polarization and reddening are both large
and incompatible with the 188 pc Reiz/Franco distance, so we chose the
Neckel et al distance. 

	These two huge disrepancies, and particularly the one for
HD100198, are illustrative of the generic problems with photometric distances. 
{\it Caveat emptor!}

\section {COMBINING THE POLARIZATIONS}

 	Whenever a star was listed more than once we took a weighted average
of the Stokes parameters in the different catalogs.  The weights were
equal to the reciprocals of the squares of the uncertainties in
polarization percentage.  Most catalogs list the formal uncertainty for
individual stars.  However, neither MFKNK nor Goodman list uncertainties
for individual stars.  In MFKNK there are different original sources and
we assigned uncertainties in percent polarization $\Delta p$ as in Table
1; for Goodman, we adopted $\Delta p = 0.1 \%$. 

	We list the percent polarization and position angle, and their
uncertainties, as derived from the weighted average of the Stokes
parameters.  The definition of these uncertainties is slightly arbitrary
for the following reason.  In principle, the errors in Stokes $Q$ and
$U$ propagate into the errors in the final polarization percentage and
position angle; Schroder provides the relevant equations.  However, if
one follows the procedure that we did---namely, to calculate $Q$ and $U$
from the polarization percentage and angle, average the Stokes
parameters, and obtain the new average polarization percentage and
angle---then the errors in the final average depend on the position
angle of the original results.  That is, the calculated uncertainty in
the final result will generally be different if one uses different zero
points for the definition of position angle, for example using Galactic
versus Equatorial position angles.  This is clearly unacceptable.  

	We solved this problem by using a modified version of the proper
formulae, as follows:

\begin{mathletters}
\begin{equation}
pp = \sqrt{ \langle Q \rangle^2 + \langle U \rangle^2 }
\end{equation}
\begin{equation}
\theta = 0.5 \ \rm{atan}\left( {\langle U \rangle \over 
     \langle Q \rangle} \right)
\end{equation}
\begin{equation}
\sigma (pp) = \sqrt{( \sigma(\langle Q \rangle)^2 + \sigma(\langle U \rangle)^2}
\end{equation}
\begin{equation}
\sigma(\theta) = {\rm atan}\left( {0.5 \sigma (pp) \over pp} \right)
\end{equation}
\end{mathletters}

\noindent In these equations, $\langle Q \rangle$ is the weighted
average of $Q$ as defined above.  Also, $\sigma(\langle Q \rangle)^2$ is
the weighted average of the squares of the residuals $(Q_i - \langle Q
\rangle)^2$, where the subscript $i$ represents the different
measurements and again the weighted average is defined above. 

	The polarization uncertainties given in the catalogs are not
always consistent with results when comparing one catalog with another. 
Schroeder (1976) provides an illustrative graphical summary of the
comparisons between his results and others.  There are significant
random differences and also systematic differences in polarization
percentage.  Angles are better defined except for some cases where there
are extremely large differences. 

	{\it Generally speaking, one should be cautious}: unless
polarizations are large, systematic errors and perhaps underquoted
random errors may make results appear more reliable than they really
are. 

\section{DISTANCES}

	We compared the distances in various catalogs; most distances
were consistent to within, say, 20\%.  However, there were some that
were highly discrepant. We present an illustrative example.  

	All catalogs except for Leroy's used spectroscopic parallax.  We
selected a set of the worst-agreeing distances and investigated them in
SIMBAD.  For example, for HD 63964 Axon and Ellis (1976) list 45 pc,
MFKNK list 40 pc, while Krautter lists 1640 pc.  SIMBAD says this is an
F5Ib star, with VMAG 8.2; without extinction, its distance modulus is
12.8 mag, so its distance is 3600 pc if there is no extinction. 
Clearly, Krautter's larger distance is correct. 

	Rather than take averages of distances, we generated a priority
list (Table 2) based in part on the amount of information given in the
original polarization catalogs; the more information (such as
extinction), the higher the priority.  We updated the spectral type,
distance, and reddening where possible; where not, we did whatever we
could.  Apart from Reiz and Franco (1998), we assigned the highest
priority to Neckel, Klare, and Sarcander (1980; CDS catalog II/62); this
catalog is devoted exclusively to distances and extinctions and they
seem to have taken great care. 

	All distances are photometric and correspondingly uncertain
because they depend on accurate spectral classification.  Errors in
distance are sometimes larger than a factor of ten.  See the discussion
above of the Reiz and Franco (1998) catalog!

\section{DESCRIPTION OF AGGLOMERATION FILE}

	Table 3 provides a byte-by-byte description of the agglomeration
file.  It contains 9286 entries, sorted in order of declination and
right ascension as embodied in $DECRA$. 

	We provide the following specific comments and cautions:

	{\bf (1)} If HDNR, BDNR, CDNR, and CPDNR all equal --999, then we
are relying on the stellar position as given in the original catalog. 
There is a possibility that either the stellar identification or the
position is incorrect. 

	{\bf (2)} $IDCAT = [1,2,3,4,5]$ means that the primary stellar
database is [Hipparcos, Tycho, SAO, SIMBAD (arcsec accuracy), SIMBAD
($\sim$ arcmin accuracy)], respectively.  Also, see section 2.  $IDCAT =
-999$ means that that the stellar position and/or the identification may
be incorrect.  $IDCAT=-998$ means that we happened to notice a more
serious problem such as a close pair of stars; we noticed only two such
entries, HD138917 and HD232588, but there might be many more. 

	{\bf (3)} If the position came from SIMBAD, denoted by $IDCAT=4$
or 5, then there is a small possibility of the position being incorrect
because of typographical error.  $IDCAT=4$ means that SIMBAD provided
positions to arcsecond accuracy or better; $IDCAT=5$ means that SIMBAD's
accuracy was less, more like an arcminute. 

	{\bf (4)} The polarization uncertainties are $\Delta pp$ and
$\Delta \theta$.  The listed values are zero when we combined individual
values that happened to be equal, because our equation (1) does not
include the measurement uncertainties in the original catalog.  This is
incorrect.  However, because the individual values {\it are} equal, such
uncertainties are likely to be small.  Negative uncertainties (--999.9,
-99.9) occur for Goodman's (1997) entries because they were
unspecified.

	{\bf (5)} $\theta_{diff}$ is the discrepancy between the position
angles in Galactic and Equatorial coordinates in the original catalog. 
It should be small.  Large values indicate a typographical error in the
original polarization catalog, and we do not know whether the equatorial
or Galactic position angle was given correctly.  Thirteen stars have
$\theta_{diff} > 10^\circ$. 

	{\bf (6)} Some E(B--V)'s are negative and were so listed in the
original catalog. 

	{\bf (7)} Distances are highly uncertain. See the discussion in
sections 3.11 and 5.

	{\bf (8)} Spectral types were taken from the original catalog
without checking.  For catalogs that were scanned, errors in scanning
may produce nonsensical entries.  Use all spectral types with great
caution!

	{\bf (9)} Values of --999.9 or --99.9 mean that the parameter was
not given in the original catalog. 

\acknowledgements

	We thank Alyssa Goodman for providing her unpublished
polarization measurements in electronic form.  It is a pleasure to thank
Adam Krigel and Doug Finkbeiner for invaluable assistance scanning the
original paper-published articles.  We made extensive use of the SIMBAD
database, and are grateful to those people who began and maintain it.

\begin{deluxetable}{ccc} 
\footnotesize
\tablecaption{ADOPTED UNCERTAINTIES IN PERCENTAGE POLARIZATION FOR MFKNK}
\label{poltable}
\tablewidth{450pt}
\tablehead{
\colhead{DISTCAT} & \colhead{REFERENCE AUTHOR} & \colhead{$\Delta p$}
}
\startdata
   1  &    Unidentified   &   0.1     \nl
   2  &     Appenzeller (1966)     &     0.032    \nl
   3  &     Appenzeller (1968)       &  0.032    \nl
   4  &     Behr (1959)      &   0.12    \nl
   5  &     Hall (1958)      &   0.20    \nl
   6  &     van Smith (1956)   &   0.40    \nl
   7  &     Schmidt (1968)      &   0.10    \nl
   8  &     Hiltner (1956)      &   0.18    \nl
   9  &     Klare and Neckel (1977)   &   0.10    \nl
   10  &     Mathewson and Ford (1970)    &  larger of $(3.5\% \ {\rm or} \ 1.1\% \times 10^{0.2VMAG})$    \nl
\enddata
\end{deluxetable}

\begin{deluxetable}{cc} 
\footnotesize
\tablecaption{PRIORITY OF DISTANCES}
\tablewidth{450pt}
\tablehead{
\colhead{NAME OF CATALOG} & \colhead{DISTCAT}
}
\startdata
Neckel et al (1980) & 120 \nl
Reiz and Franco (1998) & 130 \nl
Klare/Neckel in MFKNK & 9 \nl
MF in MFKNK  & 10 \nl
Krautter (1980)  & 40 \nl
Schmidt in MFKNK & 7 \nl
Appenzeller in MFKNK & 2, 3 \nl
Behr in MFKNK & 4 \nl
Hall in MFKNK & 5 \nl
Hiltner in MFKNK & 8 \nl
van Smith in MFKNK & 6 \nl
MFKNK/other & 1 \nl
Schroeder (1976) & 60 \nl
Markkanen (1979) & 50 \nl
Leroy (1993) & 90 \nl
Berdyugin et al (1995) & 20 \nl
Bel et al (1993) & 100 \nl
Korhonen and Reiz (1986) & 30 \nl
Axon and Ellis (1976) & 110 \nl
\enddata
\end{deluxetable}

\begin{deluxetable}{rrrl} 
\footnotesize
\tablecaption{BYTE-BY-BYTE DESCRIPTION OF AGGLOMERATION FILE$^a$}
\tablewidth{450pt}
\tablehead{
\colhead{BYTES} & \colhead{FORMAT} & \colhead{LABEL} & \colhead{EXPLANATION}
}
\startdata
1-18    &  F18.8  &  $DECRA$   &   $(\delta. alpha)_{2000}^b$  \nl
19-28   &  F10.1  &  HDNR   &   HD number  \nl
29-39   &  F11.6  &  BDNR   &   Bonner DM number  \nl
40-50   &  F11.6  &  CDNR   &   Cordoba DM number  \nl
51-61   &  F11.6  &  CPDNR  &   Cape DM number  \nl
62-70   &  F9.3   &  $pp$      &   percentage polarization  \nl
71-79   &  F9.3   &  $\Delta pp$  &  $1\sigma$ uncertainty in $pp$  \nl
80-86   &  F7.1   &  $\theta_{eq}$ & position angle, equatorial, deg  \nl
87-93   &  F7.1   &  $\Delta \theta$ & $1\sigma$ uncertainty in position angle  \nl
94-100  &  F7.1   &  $\theta_{Gal}$ & position angle, Galactic  \nl
101-109  & F9.4    &  $\ell$       &  Galactic longitude, degrees  \nl
110-118  & F9.4    &  $b$          &  Galactic latitude  \nl
119-125  &  F7.2   & E(B--V)       &  reddening, mag  \nl
126-132  &  F7.1   & $\theta_{diff}$ & discrepancy between $\theta_{eq}$
and $\theta_{gal}$  \nl
133-137  &  I5     & $IDCAT$    &  Primary stellar database  \nl
138-144  &  F7.1   & VMAG      &  visual magnitude \nl
145-152  &  f8.1   & DISTANCE   &  distance, pc \nl
155-170  &  A16    & SP         &  spectral type  \nl
171-192  &  22I1   & POLREFS    &  polarization catalog numbers$^c$  \nl
193-197  &  I5     & DISTCAT      &  distance catalog  \nl
\enddata

\tablenotetext{a}{Many items in this table are very uncertain. See the
discussion in section 6.}

\tablenotetext{b}{The declination, in units of $10^{-4}$ decimal
degrees, is to the left of the decimal point; the right ascension, in
units of $10^{-6}$ decimal hours, is to the right of the decimal point. 
For example, $ -123456.05432100$ is $\delta = -12.3456^\circ$, $\alpha =
05.4321^h$.}

\tablenotetext{c}{POLREFS is 22 binary-like numbers, with nonzero 
meaning a particular
catalog was used and zero meaning it was not. Let us define $n_b$ as the 
particular binary number, where $n_b = 0 \rightarrow 21$; each binary
number occupies byte $171 + n_b$. Each $n_b$ corresponds to a particular
reference, as follows:
  0, no entries;
  1, no entries;
  2, Reiz and Franco (1998);
  3, Bel et al (1993);
  4, Leroy (1993);
  5, Goodman (1997);
  6, Appenzeller (1974);
  7, Schroeder (1976);
  8, Markkannen (1979);
  9, Krautter (1980);
  10, Korhonen and Reiz (1986);
  11, Berdyugin et al (1995);
  12-21, the 10 different references within MFKNK. For $(n_b
= 12 \rightarrow 21)$, we have $(n_b = 22 - DISTCAT)$, 
where $(1 \le DISTCAT \le 10)$ and is listed in Tables 1 and 2. For example,
Behr in MFKNK has DISTCAT = 4, so it is represented by a 1 in
position 18. All entries are 0 or 1 except for Leroy (1993), 
for which 7, 8, 9 mean that original data are from ME, TI,
and MA, respectively, as listed in Leroy's Table 1; and for Markkannen
(1979), for which 2 and 3 specify the original data source
as given in Markkanen's Table 2.}

\end{deluxetable}


\begin{references}

\reference {} Appenzeller I. 1966, Z. Astrophysik, 64, 269

\reference {} Appenzeller, I. 1968, ApJ, 151, 907

\reference {} Appenzeller, I. 1974, A\&A, 36, 99

\reference {} Axon, D.J. and Ellis, R.S. 1976, MNRAS, 177, 499

\reference {} Behr A., 1959, Veroeff. Univ. Sternw. Goettingen N.126

\reference {} Bel, N., Lafon, J.-P.J., and Leroy, J.L. 1993, \aap, 270, 444.

\reference {} Berdyugin, A., Snare, M.-O., Teerikorpi, P. 1995, A\&A, 294, 568

\reference {} Goodman, A. 1997, private communication

\reference {} Hall J.S., 1958, Publ. U.S. Naval Obs. 2nd Ser. 17, VI

\reference {} Heiles, C. 1997, \apjs, 111, 245

\reference {} Hiltner W.A. 1956, ApJ Suppl, 2, 389

\reference {} Jenkins, L.F. 1963, General Catalogue of Trigonometric
Stellar Parallaxes, New Haven: Yale University Observatory. 

\reference {} Klare, G. and Neckel, Th. 1977, A\&AS, 27, 215

\reference {} Korhonen, T. and Reiz,A. 1986, \aaps, 64, 487

\reference {} Krautter, J. 1980, \aaps, 39, 167

\reference {} Leroy, J.L. 1993, \aap, 274, 203.

\reference {} Markkanen, T. 1979, \aap, 74, 201

\reference {} Mathewson, D.S., Ford, V.I., 1970, MmRAS, 74, 139

\reference {} Mathewson, D.S., Ford, V.I., Klare, G., Neckel, Th.,
Krautter, J. 1978, BICDS, 14, 115

\reference {} Neckel, Th., Klare, G., and Sarcander, M. 1980, \aaps, 42, 251

\reference {} Reiz, A., and Franco, G.A.P. 1998, \aaps, 130, 133

\reference {} Schmidt T. 1968, Z. Astrophysik, 68, 380

\reference {} Schroeder, R. 1976, \aaps, 23, 125

\reference {} van Smith E.P. 1956, ApJ, 124, 43

\end{references}
\end{document}